\documentclass[twocolumn,preprintnumbers,amsmath,amssymb,showpacs]{revtex4}
\usepackage{epsfig}
\usepackage{color}
\begin{document}
\setlength{\voffset}{1.0cm}
\title{Competing mechanisms of chiral symmetry breaking \\ in a generalized Gross-Neveu model}
\author{Christian Boehmer}
\author{Michael Thies\footnote{thies@theorie3.physik.uni-erlangen.de}}
\affiliation{Institut f\"ur Theoretische Physik III,
Universit\"at Erlangen-N\"urnberg, D-91058 Erlangen, Germany}
\date{\today}
\begin{abstract}
Chiral symmetry of the 2-dimensional chiral Gross-Neveu model is broken explicitly by a bare mass term as well as a splitting of scalar and 
pseudo-scalar coupling constants. The vacuum and light hadrons --- mesons and baryons which become massless in the chiral limit ---
are explored analytically in leading order of the derivative expansion by means of a double sine-Gordon equation.
Depending on the parameters, this model features new phenomena as compared to previously investigated 4-fermion models:
spontaneous breaking of parity, a non-trivial chiral vacuum angle, twisted kink-like baryons whose baryon number reflects the
vacuum angle, crystals with alternating baryons, and appearance of a false vacuum.
\end{abstract}
\pacs{11.10.Kk,11.10.St,11.30.Cp}
\maketitle

Consider the Lagrangian density of $N$ species of massive Dirac fermions in 1+1 dimensions with attractive, U($N$) invariant  scalar
and pseudoscalar interactions,
\begin{equation}
{\cal L}  =  \bar{\psi} \left( {\rm i} \gamma^{\mu} \partial_{\mu}-m_0 \right) \psi + \frac{g^2}{2} (\bar{\psi}\psi)^2
+ \frac{G^2}{2}  (\bar{\psi}{\rm i}\gamma_5 \psi)^2.
\label{A1}
\end{equation}
Flavor indices are suppressed ($\bar{\psi}\psi = \sum_{k=1}^N \bar{\psi}_k \psi_k$ etc.) and the large $N$ limit will be assumed. This 
3-parameter field theoretic model generalizes the (massive) chiral Gross-Neveu (GN) model \cite{1} to two different coupling constants. 
Its massless 2-parameter version is related to the early work of Klimenko \cite{1a} and has only recently been investigated 
comprehensively \cite{2}. Our main motivation for considering
the Lagrangian (\ref{A1}) is to study the 
competition of two different mechanisms of explicit chiral symmetry breaking, both of which are well understood in isolation. The first 
one is kinematical and familiar from gauge theories ---  the bare fermion mass. The second one is dynamical --- breaking chiral 
symmetry through the interaction term while preserving parity. This seems to have no analogue in pure gauge theories. 
In the present work we do not attempt a complete solution of the model (\ref{A1}) which would require extensive numerical computations.
To get a first overview of its physics content, we focus on the vicinity of the chiral limit at zero temperature, where everything can be 
done in closed analytical form.

Following 't~Hooft \cite{3}, the large $N$ limit is implemented by letting $N\to \infty$ while keeping $Ng^2$ and $NG^2$ constant. As is well 
known,
this justifies the use of semiclassical methods \cite{1,4}. Thereby the  Euler-Lagrange equation of the Lagrangian (\ref{A1}) gets converted
 into the
Dirac-Hartree-Fock equation, 
\begin{equation}
\left({\rm i} \gamma^{\mu} \partial_{\mu} -S-{\rm i}\gamma_5 P\right)\psi=0,
\label{A2}
\end{equation}
where the scalar and pseudo-scalar mean fields are related to condensates (ground state expectation values) through
\begin{eqnarray}
S & = & - g^2 \langle \bar{\psi}\psi \rangle + m_0,
\nonumber \\
P & = & - G^2 \langle \bar{\psi} {\rm i} \gamma_5 \psi \rangle.
\label{A3}
\end{eqnarray}
Further simplifications arise if we concentrate on static problems in the vicinity of the chiral limit, where the potentials are slowly varying in 
space.
This allows us to invoke a systematic expansion in derivatives of $S$ and $P$ without assuming that the potentials are weak \cite{5,6}.
As a result, we arrive at an effective bosonic field theory in which the Hartree-Fock potentials appear as complex scalar field 
(written here in polar coordinates),
\begin{equation}
S-{\rm i}P = \rho {\rm e}^{{\rm i} \theta}.
\label{A4}
\end{equation}
Note that this method can only handle full occupation of single particle levels at present. It was pioneered in Ref.~\cite{7} and applied
systematically to two variants of the  Lagrangian (\ref{A1}), the massive chiral GN model ($g^2=G^2$, Ref.~\cite{6}) and the massless
generalized GN model ($m_0=0$, Ref.~\cite{2}). Since
 the form of the 
Hartree-Fock equation is the same in all of these cases, the problem at hand differs from previous ones only through the form of the double
 counting correction to the energy density,
\begin{eqnarray}
{\cal E}_{\rm d.c.} &=& \frac{(S-m_0)^2}{2 Ng^2}+\frac{P^2}{2 N G^2}
\nonumber \\
& = & \frac{\rho^2 \cos^2 \theta}{2Ng^2} - \frac{m_0 \rho \cos \theta}{Ng^2}+ \frac{\rho^2 \sin^2 \theta}{2NG^2}.
\label{A5}
\end{eqnarray}  
An irrelevant term $\sim m_0^2$ has been dropped. Regularization and renormalization require only a straightforward extension 
of previous works. We replace the 3 bare parameters ($m_0,g^2,G^2$) by physical parameters 
($\xi_1,\xi_2,\eta$) via
\begin{eqnarray}
\frac{\pi}{Ng^2} & = & \ln \Lambda + \xi_1,
\nonumber \\
\frac{\pi}{NG^2} & = &  \ln \Lambda + \xi_2,
\nonumber \\
\frac{\pi m_0}{Ng^2} & = & \eta.
\label{A6}
\end{eqnarray}
The $\ln \Lambda$ dependence is mandatory to ensure that the ultraviolet divergence in the sum over single particle energies is cancelled 
by the double counting correction. In the last line of Eq.~(\ref{A6}), we avoid the use of the standard confinement parameter \cite{8}
\begin{equation}
\gamma = \frac{\pi m_0}{Ng^2 m} = \frac{\eta}{m}
\label{A7}
\end{equation}
at this stage. This is done in order not to mix the parameters of the model with dynamical quantities, which may lead to confusion in
the present 3-parameter model. Restricting ourselves to the leading order of the 
derivative
expansion, we assume furthermore that the radius $\rho$ is fixed
at the dynamical fermion mass and that the chiral angle field $\theta$ is slowly varying. These assumptions can be justified by looking at 
higher order terms
of the derivative expansion, but they also have a very simple physical basis: Close to the chiral limit, the would-be Goldstone 
field $\theta$ 
(the ``pion" field) is the only one which can be modulated at low cost of energy \cite{7}. The renormalized ground state energy density 
(including the vacuum contribution) corresponding to Lagrangian (\ref{A1}) then reads
\begin{eqnarray}
2 \pi {\cal E} &=& \rho^2 \left( \ln \rho - \frac{1}{2} \right) +  \frac{1}{4} \rho^2 (\theta')^2 - 2 \eta \rho \cos \theta
\nonumber \\
& & + \xi_1 \rho^2 \cos^2 \theta + \xi_2 \rho^2 \sin^2 \theta .
\label{A8}
\end{eqnarray}
Fermion number is given by the winding number of the chiral field \cite{6}
\begin{equation}
N_f = \frac{N}{2\pi}  \int_{-\infty}^{\infty} {\rm d}x \theta' = \frac{N}{2\pi} \left[ \theta(\infty)- \theta(-\infty) \right].
\label{A9}
\end{equation}
All we have to do is to minimize the energy $\int {\rm d}x {\cal E}$
classically. As a result, we will get information on the vacuum and its symmetries, as well as on light mesons
and baryons in the vicinity of the chiral limit. For a homogeneous vacuum, the truncated derivative 
expansion is exact 
since the condensates are spatially constant. Hence our results for the vacuum may be taken as the large $N$ limit without
any further approximation. Light hadrons are those which become massless in the chiral limit. Here the derivative expansion
can be viewed as a kind of chiral perturbation
theory, reliable close to the chiral limit. The expression for the pion mass for example is of the type of the Gell-Mann, Oakes, 
Renner (GOR) 
relation \cite{9} in the real world. The fact that baryons emerge from a non-linear theory for the pion field with the baryon 
number as topological winding number is of course reminiscent of the Skyrme model in 3+1 dimensions \cite{10,11}.

We first determine the vacuum. To this end, we minimize $2\pi {\cal E}$, Eq.~(\ref{A8}), with respect to ($x$-independent) $\rho$ and $\theta$ 
-- the dynamical fermion mass and chiral vacuum angle. This yields the transcendental equations
\begin{eqnarray}
0 & = & \ln \rho +\xi_1 \cos^2 \theta  +\xi_2 \sin^2 \theta - \frac{\eta}{\rho} \cos \theta,
\nonumber \\
0 & = & \sin \theta \left(  \cos \theta - \frac{\eta}{\rho(\xi_1-\xi_2)}\right).
\label{A10}
\end{eqnarray}
Their solution requires a case differentiation. To understand qualitatively what
to expect, let us temporarily choose units such that the dynamical fermion mass is 1 ($\rho=1$) and focus on the $\theta$-dependent
part of the vacuum energy density,
\begin{equation}
2\pi \tilde{\cal E}(\theta) = -2\gamma \cos \theta - \frac{1}{2} \xi \cos (2\theta), \qquad \xi=\xi_2-\xi_1.
\label{A11}
\end{equation}
We have used the fact that the distinction between $\eta$ and the confinement parameter $\gamma$ disappears in these units, cf.
Eq.~(\ref{A7}). Note also that depending on the sign of $\xi$, either the scalar coupling (for $\xi>0$) or the pseudoscalar coupling (for $\xi<0$)
dominates.

\begin{figure}
\begin{center}
\epsfig{file=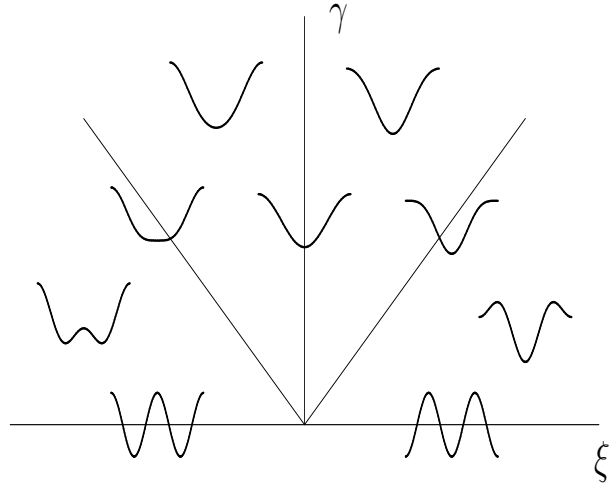,height=8cm,width=6.4cm,angle=270}
\caption{Qualitative shapes of effective potentials, Eq.~(\ref{A11}), in the ($\xi,\gamma$) half plane. Each inserted plot shows
$\tilde{\cal E}(\theta)$ in the interval $[-\pi,\pi]$, so that the endpoints have to be identified. The origin ($\xi=0,\gamma=0$) is the chirally
symmetric point where the effective potential vanishes. When crossing the critical lines $\gamma=\pm \xi$, the number
of extrema changes.} 
\label{fig1}
\end{center}
\end{figure}
A survey of the $\theta$-dependence of this effective potential in the ($\xi, \gamma$) half plane ($\gamma \geq 0$) reveals 
a rich landscape (see Fig.~\ref{fig1}): At the origin ($\xi=0,\gamma=0$), the potential is identically zero (not shown in Fig.~1) and the
vacuum infinitely degenerate. This is the U(1) chirally symmetric point. 
Along the $\gamma$ axis there is a minimum at $\theta=0$ and a maximum at $\theta=\pi$ --- the massive chiral GN model. Along 
the $\xi$ axis,
there are two degenerate minima separated by two degenerate maxima --- the massless generalized GN model. As discussed in Ref.~\cite{2},
the minima can be identified with $0$ and $\pi$ for both $\xi>0$ and $\xi<0$ by means of a global chiral rotation, so that the positive and 
negative $\xi$ half-axes are in fact equivalent. What happens in the parameter region away from the $\gamma$- and $\xi$-axes
depends on the sign of $\xi$. If $\xi>0$, the quadratic maximum becomes a quartic maximum
at $\gamma=\xi$; for larger values of $\xi$, a false vacuum develops at $\theta=\pi$. In the limit $\gamma \to 0$
the two minima become degenerate. If $\xi<0$ on the other hand, the quadratic minimum becomes quartic when crossing the critical 
line $\gamma=-\xi$. This is indicative of a pitchfork bifurcation with two symmetric, degenerate minima present for $\gamma < -\xi$.
A non-trivial vacuum angle signals a non-vanishing pseudoscalar condensate and hence a breakdown of parity. 
This breakdown of parity is spontaneous, but induced by the explicit breaking of chiral symmetry.

With this overall picture in mind, we return to Eqs.~(\ref{A10}) and solve them in  two distinct cases: 
\begin{itemize}
\item Unbroken parity (phase I)
\begin{eqnarray}
\theta_{\rm vac} & = & 0
\nonumber \\
\rho_{\rm vac} & = & \frac{\eta}{W(\eta {\rm e}^{\xi_1})},
\nonumber \\
2 \pi {\cal E}_{\rm vac} & = & - \frac{\eta^2}{2} \left( \frac{1 + 2 W(\eta {\rm e}^{\xi_1})}{W^2(\eta {\rm e}^{\xi_1})}\right).
\label{A12}
\end{eqnarray}
\item Broken parity (phase II)
\begin{eqnarray}
\theta_{\rm vac} & = & \pm \arccos \frac{\eta {\rm e}^{\xi_2}}{\xi_1-\xi_2}, 
\nonumber \\
\rho_{\rm vac} & = & {\rm e}^{- \xi_2},
\nonumber \\
2 \pi {\cal E}_{\rm vac} & = & - \frac{1}{2} {\rm e}^{-2\xi_2}- \frac{\eta^2}{\xi_1-\xi_2}.
\label{A13}
\end{eqnarray}
\end{itemize}
In Eqs.~(\ref{A12}) we have introduced the Lambert $W$ function with the defining property
\begin{equation}
x=W(x){\rm e}^{W(x)}.
\label{A14}
\end{equation}
The vacuum energy in the parity broken phase II is lower than in the symmetric phase I. However, phase II only exists
if $\theta_{\rm vac}$ is real or, equivalently,
\begin{equation}
\xi_1-\xi_2 \geq W(\eta {\rm e}^{\xi_1}).
\label{A15}
\end{equation}

The next steps can be further simplified as follows. After minimization and determining the phase on the basis
of Eq.~(\ref{A15}), we normalize the radius of the chiral circle (the physical fermion mass) to 1 by a choice of units,
\begin{equation}
\rho=\rho_{\rm vac} = 1.
\label{A16}
\end{equation} 
Then $\eta$ may be identified with the confinement parameter (\ref{A7}) familiar from the standard massive GN models,
\begin{equation}
\eta = \rho \gamma \to  \gamma.
\label{A17}
\end{equation}
In phase I, the condition $\rho=1$ implies
\begin{equation}
\xi_1 = \gamma.
\label{A18}
\end{equation}
The vacuum energy density becomes
\begin{equation}
 {\cal E}_{\rm vac}^I = - \frac{1}{4\pi} - \frac{\gamma}{2\pi},
\label{A19}
\end{equation}
in agreement with the standard massive GN models.
The $\theta$-dependent part of the energy density will be needed for the analysis of light mesons and baryons; in phase I 
it is given by
\begin{equation}
2\pi {\cal E}_{\theta}^I = \frac{1}{4} (\theta')^2 - 2 \gamma \cos \theta -\frac{1}{2} (\xi_2 - \gamma) \cos (2\theta). 
\label{A20}
\end{equation}
In phase II, the condition $\rho=1$ implies
\begin{equation}
\xi_2 = 0, 
\label{A21}
\end{equation}
whereas the vacuum energy density assumes the form
\begin{equation}
{\cal E}_{\rm vac}^{II}=-\frac{1}{4\pi} - \frac{\gamma^2}{2\pi \xi_1}.
\label{A22}
\end{equation}
In this phase, the $\theta$-dependent part of the energy density reads
\begin{equation}
2\pi {\cal E}_{\theta}^{II} = \frac{1}{4} (\theta')^2 - 2 \gamma \cos \theta +\frac{1}{2} \xi_1 \cos (2\theta) .
\label{A23}
\end{equation}
Eqs.~(\ref{A20},\ref{A23}) can be treated simultaneously  by setting
\begin{equation}
2\pi {\cal E}_{\theta} = \frac{1}{4} (\theta')^2 - 2 \gamma \cos \theta - \frac{1}{2} \xi \cos (2\theta) 
\label{A24}
\end{equation}
with the definition
\begin{equation}
\xi  =  \xi_2 - \xi_1   = \left\{ 
\begin{array}{ll}  \xi_2- \gamma & ({\rm phase}\ I, \xi>-\gamma) \\
 - \xi_1 & ({\rm phase}\ II, \xi<-\gamma) \end{array} \right.
\label{A25}
\end{equation}
We have traded the original bare parameters $g^2, G^2, \eta$ against two dimensionless parameters $\gamma, \xi$ and one scale,
the dynamical fermion mass $\rho=1$. The notation is chosen so as to agree with previous results for the massive chiral GN model
\cite{6} for $\xi=0$ and the massless generalized GN model \cite{2} for $\gamma=0$.

Next consider the light meson mass in both phases. Expanding expression (\ref{A24}) around the vacuum angle $\theta_{\rm vac}$ to 2nd
order in $\vartheta= \theta-\theta_{\rm vac}$, we can simply read off the pion mass as follows:
\begin{itemize}
\item Phase I ($\xi>-\gamma$)
\begin{eqnarray}
\theta_{\rm vac} & = & 0
\nonumber \\
2\pi {\cal E} & \approx & \frac{1}{4} (\vartheta')^2 + (\gamma+\xi)\vartheta^2 + {\rm const.}
\nonumber \\
m_{\pi}^2 & = & 4 (\gamma+\xi)
\label{A26}
\end{eqnarray}
\item Phase II ($\xi<-\gamma$)
\begin{eqnarray}
\theta_{\rm vac} & = & \pm \arccos \left( - \frac{\gamma}{\xi}\right) \ = \ \pm 2 \arctan \sqrt{\frac{\xi + \gamma}{\xi - \gamma}}
\nonumber \\
2\pi {\cal E} & \approx & \frac{1}{4} (\vartheta')^2 + \left( \frac{\gamma^2-\xi^2}{\xi}\right) \vartheta^2 + {\rm const.}
\nonumber \\
m_{\pi}^2 & = & 4 \left(\frac{\gamma^2-\xi^2}{\xi}\right)
\label{A27}
\end{eqnarray}
\end{itemize}
The last lines of Eqs.~(\ref{A26},\ref{A27}) may be regarded as the generalized GOR relations in our model. 

\begin{figure}
\begin{center}
\epsfig{file=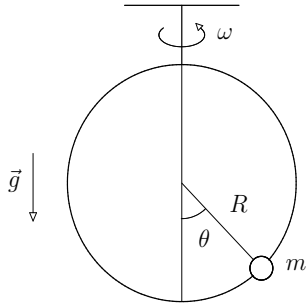,height=4.0cm,width=4.0cm,angle=270}
\caption{Mechanical model illustrating vacuum structure, symmetries and meson masses of the generalized massive GN model
($\xi<0$) in the vicinity of the chiral limit.}
\label{fig2}
\end{center}
\end{figure}
It is amusing that a well-known mechanical system is closely analogue to the present problem in the case $\xi<0$, cf. Fig.~\ref{fig2}:
A bead (mass $m$) is sliding without friction on a circular hoop (radius $R$) in a homogeneous gravitational field. 
The hoop rotates with constant  angular velocity $\omega$ around a vertical axis through its center. The Lagrangian reads
\begin{equation}
L = \frac{1}{2} m R^2 \left( \dot{\theta}^2+ \omega^2 \sin^2 \theta\right) + mgR\cos \theta.
\label{A28}
\end{equation}
Denote the pendulum frequency by $\omega_0=\sqrt{g/R}$. At $\omega=0$, there is 
a unique stable minimum at $\theta=0$, accompanied by small oscillations of frequency $\omega_0$. If one increases $\omega$,
this minimum stays stable at first, but the frequency decreases like $\sqrt{\omega_0^2-\omega^2}$ until it vanishes at the critical value
$\omega=\omega_0$. At this point, two symmetric stable minima at $\theta=\pm {\rm arccos}\,( \omega_0^2/\omega^2)$
develop, a textbook example of a pitchfork bifurcation \cite{12}. Beyond this point, the frequency of small oscillations is replaced by 
$\sqrt{\omega^4-\omega_0^4}/\omega$. The gravitational field and the uniform rotation are two distinct  mechanisms of breaking
the original SO(2) symmetry of the circle. The mapping of this mechanical problem onto our field theory model is obvious:
U(1) chiral symmetry corresponds to the rotational symmetry of the circle, the bare mass plays the role of gravity, the difference
in coupling constants corresponds to the uniform rotation, the pion masses to the frequencies of small oscillations. We only have to
identify $ \gamma=\omega_0^2/4, \xi=- \omega^2/4$ to map the two problems onto each other quantitatively. In principle, the regime $\xi>0$
 could
also be modeled by assuming that the particle is charged and invoking an additional constant magnetic field, but in the absence of a phase
transition this is less instructive.

Let us now turn to baryons and baryon crystals. Here we need large amplitude solutions of the equation
\begin{equation}
\theta'' = 4 \gamma \sin \theta + 2 \xi \sin 2 \theta.
\label{A29}
\end{equation}
For small values of the parameters $\xi,\eta$ the kink-like soliton solutions of this equation are slowly varying so that the derivative
 expansion is applicable.
The same is true for periodic soliton crystal solutions at sufficiently low density. However there is no restriction on the ratio $\xi/\gamma$, so 
that the full phase structure shown in Fig.~\ref{fig1} is accessible in the vicinity of the chiral limit.
Since Eq.~(\ref{A29}) has no explicit $x$-dependence, it can be integrated once,
\begin{equation}
\frac{1}{2} (\theta')^2+ 4 \gamma \cos \theta + \xi \cos (2\theta) = {\rm const.}
\label{A30}
\end{equation}
The second integration is then carried out by separation of variables.

The mechanical  interpretation of the kinks is well-known: If we interpret $x$ as time
coordinate, Eq.~(\ref{A29}) describes motion of a classical particle in a potential inverted as compared to the potentials
shown in Fig.~\ref{fig1}. The kink-like tunneling solutions between different vacua in field theory go over into
classical paths joining two degenerate maxima in the mechanics case. In this classical mechanics interpretation, Eq.~(\ref{A30})
expresses conservation of the Hamilton
function. As a matter of fact, Eq.~(\ref{A29}) is nothing but the double sine-Gordon equation, a widely used generalization
of the sine-Gordon equation to which it reduces if either $\gamma$ or $\xi$ vanishes. Its solutions can be found in the literature, see 
e.g. \cite{13}, so that we refrain from giving any details of the derivation. Since $\theta$ is an angular variable, kinks do exist everywhere
in the ($\xi,\gamma$) half-plane. Inspection of the effective potentials of Fig.~\ref{fig1} then helps to understand the following results:

For  $\xi>-\gamma$ (phase I) there is only one kink solution
\begin{equation}
\theta_{\rm kink} = - 2 \arctan \frac{\sqrt{\xi + \gamma}}{\sqrt{\gamma} \sinh(2\sqrt{\xi+\gamma}x)}.
\label{A31}
\end{equation}
We define the branch of the arctan such that $\theta$ goes from 0 to $2\pi$ along the $x$ axis.
For $\xi<-\gamma$ (phase II) there are two different kinks depending on how one connects the minima along the chiral circle,
\begin{eqnarray}
\theta_{\rm large} & = &  - 2 \arctan \left[ \sqrt{\frac{\xi+\gamma}{\xi-\gamma}}\coth \left( \sqrt{\frac{\gamma^2-\xi^2}{\xi}}x\right)\right],
\nonumber \\
\theta_{\rm small} & = & + 2 \arctan \left[ \sqrt{\frac{\xi+\gamma}{\xi-\gamma}}\tanh \left( \sqrt{\frac{\gamma^2-\xi^2}{\xi}}x\right)\right].
\nonumber \\
\label{A32}
\end{eqnarray}
Here, our choice of the branch of arctan is such that $\theta$ goes from $- \theta_{\rm vac}$ to $\theta_{\rm vac}$ for the small kink and
from $\theta_{\rm vac}$ to $2\pi - \theta_{\rm vac}$ for the large kink. The baryon numbers $B=N_f/N$ are
\begin{eqnarray}
B_{\rm kink} & = & 1,
\nonumber \\
B_{\rm large} & = & 1- \frac{\theta_{\rm vac}}{\pi},
\nonumber \\
B_{\rm small} & = & \frac{\theta_{\rm vac}}{\pi},
\label{A33}
\end{eqnarray}
with $\theta_{\rm vac}$ from Eq.~(\ref{A27}) with the + sign. The terms small and large refer to the chiral twist of the two kinks which
in turn is reflected in the baryon number. The baryon numbers of a small and a large kink add up to 1 simply because these kinks
correspond to
the 2 possibilities of travelling from one minimum to the other one along a circle.  Eqs.~(\ref{A31}-\ref{A33}) refer to kinks with
positive baryon number. By changing the sign of $\theta$, these can be converted into
antikinks with opposite baryon number.

\begin{figure}
\begin{center}
\epsfig{file=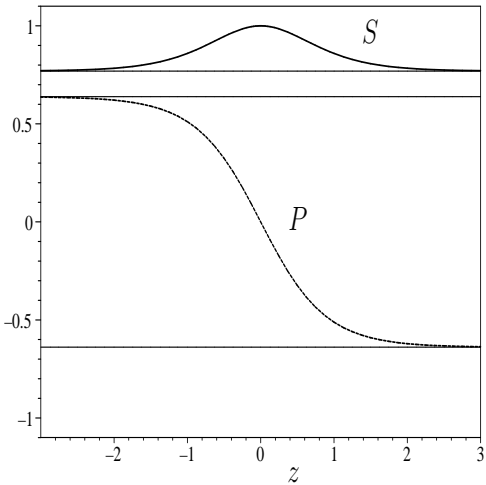,height=6.4cm,width=6.4cm,angle=270}
\caption{Scalar and pseudoscalar potentials for the small kink baryon in the parity broken phase II with $\xi=-1.3 \gamma$ as a
function of $z=m_{\pi}x$.
The straight lines are the asymptotic values coinciding with the vacuum condensates. There are two degenerate vacua with
equal scalar and opposite pseudoscalar condensates, related by a parity transformation.}
\label{fig3}
\end{center}
\end{figure}
\begin{figure}
\begin{center}
\epsfig{file=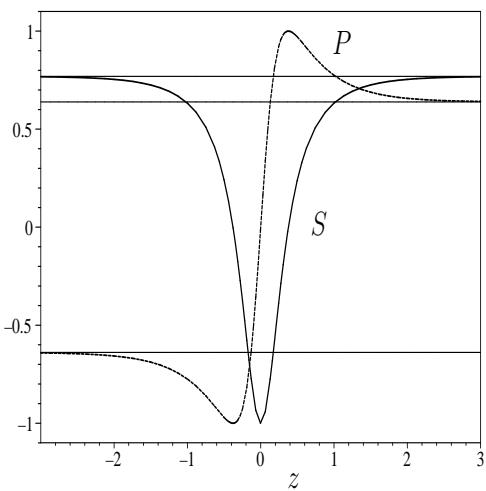,height=6.4cm,width=6.4cm,angle=270}
\caption{Same as Fig.~\ref{fig3} but for the large kink baryon.}
\label{fig4}
\end{center}
\end{figure}
\begin{figure}
\begin{center}
\epsfig{file=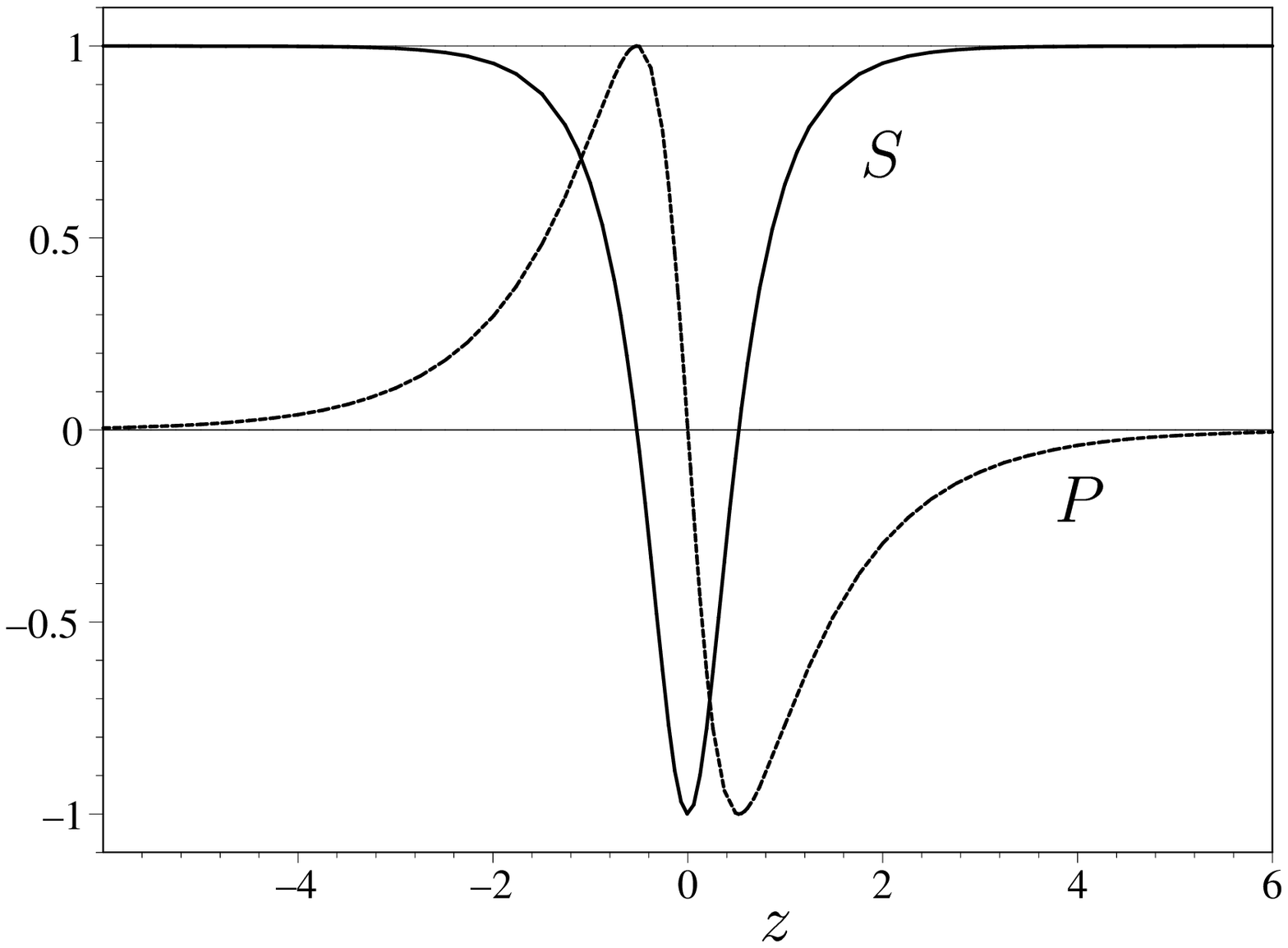,height=6.4cm,width=6.4cm,angle=270}
\caption{Same as Fig.~\ref{fig3} but for the kink baryon in the parity restored phase I and $\xi=-0.7 \gamma$. The pseudoscalar
vacuum condensate vanishes.}
\label{fig5}
\end{center}
\end{figure}
In Figs.~\ref{fig3} and \ref{fig4}, we illustrate the scalar and pseudoscalar potentials for the small and large kinks in the parity broken phase II.
To understand these graphs, we recall that the two vacua are characterized by the chiral angles $\pm \theta_{\rm vac}$, Eq.~(\ref{A27}).
The parity even, scalar vacuum condensate ($\cos \theta_{\rm vac}$) is the same in both vacua, the parity odd, pseudoscalar condensate
($-\sin \theta_{\rm vac}$) has opposite sign. This is reflected in the asymptotic behavior of $S$ and $P$ for the kinks which connect
these two vacua. To contrast this behavior with baryons in phase I (unbroken parity, $\xi>-\gamma$), we show in Fig.~\ref{fig5}
the kink baryon from Eq.~(\ref{A31}) where now both $S$ and $P$ are periodic. For the parameters chosen here, it resembles closely the
standard sine-Gordon kink.

\begin{figure}
\begin{center}
\epsfig{file=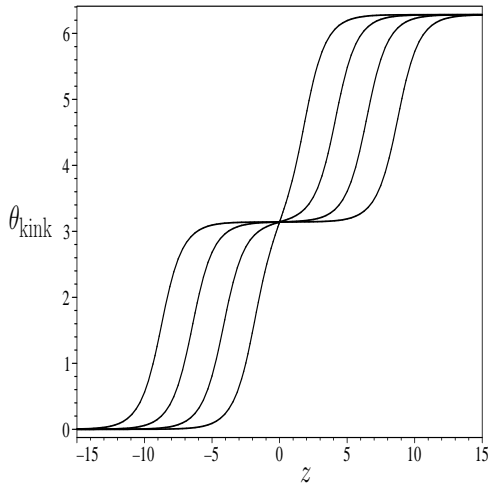,height=6.4cm,width=6.4cm,angle=270}
\caption{Behavior of $\theta_{\rm kink}$ as a function of $z=m_{\pi}x$, for $\gamma/\xi= 10^{-1},10^{-3},10^{-5},10^{-7}$, with increasing width
of the plateau for decreasing ratio $\gamma/\xi$. The plateau becomes infinitely wide in the limit $\gamma \to 0$, leading to decoupled kinks
with baryon number 1/2 in the massless model \cite{2}.}
\label{fig6}
\end{center}
\end{figure}

Note the following limits:
\begin{itemize}
\item
Massive NJL model ($\gamma>0, \xi=0$): There is a unique minimum at $\theta=0$. We recover previous (sine-Gordon) results \cite{6,7} 
with the help of the identity
\begin{equation}
\theta = \mp 2 \arctan \frac{1}{\sinh (2\sqrt{\gamma}x)} = \pm 4 \arctan {\rm e}^{2 \sqrt{\gamma}x}.
\label{A34}
\end{equation}

\item 
Massless generalized GN model ($\gamma=0, \xi>0$): There are 2 degenerate minima at $\theta=0,\pi$ and correspondingly 2 kink baryons
with baryon number 1/2. The limit is singular (see Fig.~\ref{fig6}): As $\gamma \to 0$, the kink develops a plateau which becomes infinitely wide 
at $\gamma=0$. The kink decouples into 2 half-kinks each carrying baryon number 1/2 \cite{2}. As one sees in Fig.~\ref{fig1},
this happens when the maxima in the inverted potential become degenerate or, equivalently, the false vacuum and the true vacuum
in the original potential become equal.

\end{itemize}

It is worth mentioning that there is yet another solitonic solution of some physics relevance: If one is interested in the decay
of the false vacuum, one has to consider tunneling through the barrier. This in turn is related to the kink-antikink which 
starts from the lower maximum, is reflected at the barrier and returns to the starting point (the bounce \cite{14}).
Since we are mainly interested in the vacuum and low-lying hadrons here, we do not go further into this problem.

We now turn to a useful test of the consistency of our results, following Ref.~\cite{15}. Consider the 
divergence of the axial current as obtained from the Euler-Lagrange equations for the Lagrangian (\ref{A1}),
\begin{equation}
\partial_{\mu} j_5^{\mu} = 2 \bar{\psi} {\rm i} \gamma_5 \psi \left[ m_0-(g^2-G^2)\bar{\psi}\psi \right].
\label{A35}
\end{equation}
The right-hand side exhibits the 2 sources of chiral symmetry breaking, the bare fermion mass and the splitting of the coupling
constants. The self-consistency conditions (\ref{A3}) and the renormalization scheme (\ref{A6}) can be used to rewrite Eq.~(\ref{A35}) as
\begin{equation}
\partial_{\mu} j_5^{\mu} = - \frac{2NP}{\pi}\left[ \eta - (\xi_1-\xi_2)S\right]
\label{A36}
\end{equation}
or, in units $\rho=1$ and with the notation of Eqs.~(\ref{A17},\ref{A25}),
\begin{equation}
\partial_{\mu} j_5^{\mu} = - \frac{2NP}{\pi}\left( \gamma + \xi S\right).
\label{A37}
\end{equation}
Taking the expectation value of this equation in a time-independent state and remembering that $j_5^1 = j^0$ in 1+1 dimensions, we arrive
at the following expression for the fermion density,
\begin{equation}
j^0(x)  =   - \frac{2N}{\pi} \int_{-\infty}^x {\rm d}x' P(x') \left[ \gamma + \xi S(x')\right],
\label{A38}
\end{equation}
and, after another integration, the sum rule
\begin{equation}
N_f  =   \frac{2N}{\pi} \int_{-\infty}^{\infty} {\rm d}x x P(x) \left[ \gamma + \xi S(x)\right].
\label{A39}
\end{equation}
The last equation in particular provides us with a non-trivial way of testing
the baryon potentials. By inserting $S=\cos \theta$ and $P=-\sin \theta$ into the sum rule with $\theta$ from Eqs.~(\ref{A31},\ref{A32}),
we indeed reproduce the baryon numbers (\ref{A33}). 
Notice also that the expectation value of Eq.~(\ref{A36}) for the divergence of the axial current,
\begin{equation}
(j^0)'(x) = - \frac{2NP(x)}{\pi}\left[ \gamma+ \xi S(x)\right],
\label{A40}
\end{equation}
reduces to the double sine-Gordon equation, Eq.~(\ref{A29}), if we insert
\begin{equation}
j^0(x) = \frac{N}{2\pi} \theta'(x)
\label{A41}
\end{equation}
and express $S,P$ in terms of the chiral angle $\theta$. This points to an alternative derivation of the basic equation (\ref{A29}) 
which would not even require the derivative expansion, at least to leading order considered here. 

It is straightforward to compute the baryon masses by integrating the energy density and subtracting the vacuum contribution,
\begin{eqnarray}
2\pi M & = &  \int {\rm d}x \left\{ \frac{1}{4} [\theta'(x)]^2 - 2\gamma (\cos \theta(x)-\cos \theta_{\rm vac}) \right.
\nonumber \\
& & \left. - \frac{1}{2}\xi \left[ \cos (2\theta(x))- \cos(2\theta_{\rm vac})\right] \right\}.
\label{A42}
\end{eqnarray}
One finds
\begin{eqnarray}
M_{\rm kink} & = & \frac{2\sqrt{\gamma+\xi}}{\pi} + \frac{\gamma}{\pi \sqrt{\xi}} \ln \left( \frac{\sqrt{\gamma+\xi}+\sqrt{\xi}}{\sqrt{\gamma+\xi}-
\sqrt{\xi}}\right),
\nonumber \\
M_{\rm large} & = & \frac{1}{\pi} \sqrt{\frac{\xi^2-\gamma^2}{-\xi}}+ \frac{2\gamma}{\pi \sqrt{-\xi}}\arctan \sqrt{\frac{\xi-\gamma}{\xi +\gamma}},
\nonumber \\
M_{\rm small} & = & \frac{1}{\pi} \sqrt{\frac{\xi^2-\gamma^2}{-\xi}} - \frac{2\gamma}{\pi \sqrt{-\xi}}\arctan \sqrt{\frac{\xi+\gamma}{\xi -\gamma}},
\nonumber \\
\label{A43}
\end{eqnarray} 
and the same results for the corresponding antikinks. These expressions
are of course known from studies of the classical double sine-Gordon equation.

\begin{figure}
\begin{center}
\epsfig{file=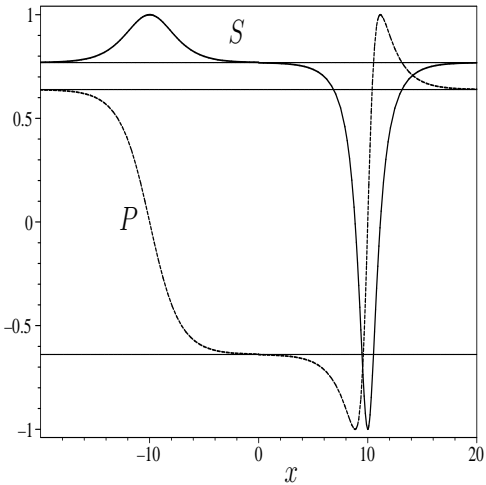,height=6.4cm,width=6.4cm,angle=270}
\caption{Unit cell of soliton crystal in the parity broken phase for $\gamma=0.2, \xi=-0.26$. Small and a large kinks must alternate 
in the crystal due to their different asymptotics. The baryon numbers of the 2 constituents in the unit cell add up to 1.}
\label{fig7}
\end{center}
\end{figure}

Finally, consider baryonic matter at low density. The pertinent solutions of the double sine-Gordon equation are kink crystals which can
be evaluated analytically in terms of Jacobi elliptic functions. Since we work only to lowest order of the derivative expansion in the present
study, we bypass the complicated exact solution by simply gluing together kink solutions. This is adequate in the low density limit.
In the parity preserving phase I, the basic building block is $\theta_{\rm kink}$, Eq.~(\ref{A31}). Let us denote the separation between two
 kinks (i.e., the lattice constant) 
by $d$, so that the baryon density is $\rho_B=1/d$. A dilute periodic array of kinks is then well approximated by
\begin{equation}
\theta_{\rm crystal}^I = \theta_{\rm kink}(x-nd)+2\pi n  \ {\rm for} \  x\in [nd-d/2,nd+d/2].
\label{A44}
\end{equation}
For sufficiently large $d$ this yields a smooth staircase curve which solves the double sine Gordon equation exactly
except at the gluing points $x=(n+1/2)d$.  There the error can be made arbitrarily small for large $d$. The energy density
in the dilute limit is just $M_{\rm kink}\rho_B$ with the kink mass from Eq.~(\ref{A43}). In phase II, we have to proceed slightly differently.
Obviously one can only glue together the small and large kinks in an alternating way, see Figs.~\ref{fig3},\ref{fig4}. We therefore
first construct a unit cell of the crystal by joining one small and one large kink,
\begin{equation}
\tilde{\theta}_{\rm kink}(x)  =  \left\{ \begin{array}{lll}
\theta_{\rm small}(x+d/4) & {\rm for} & -d/2 <x<0 \\
\theta_{\rm large}(x-d/4) & {\rm for} &  0 <x< d/2 \end{array} \right.
\label{A45}
\end{equation}
This carries baryon number 1 and is periodic modulo $2\pi$, so that the unit cells can now be assembled into a crystal in the same way as in 
phase I, Eq.~(\ref{A44}),
\begin{equation}
\theta_{\rm crystal}^{II} = \tilde{\theta}_{\rm kink}(x-nd)+2\pi n \ {\rm for} \  x\in [nd-d/2,nd+d/2].
\label{A46}
\end{equation}
The energy density in the low density limit of phase II becomes 
\begin{equation}
{\cal E} = (M_{\rm small} + M_{\rm large})\rho_B
\label{A47}
\end{equation}
where the sum of the kink masses from Eqs.~(\ref{A43}) can be simplified to 
\begin{equation}
M_{\rm small} +M_{\rm large} = \frac{2}{\pi} \sqrt{\frac{\xi^2-\gamma^2}{-\xi}} +  \frac{2\gamma}{\pi \sqrt{-\xi}}\arctan \frac{\gamma}{\sqrt{\xi^2-\gamma^2}}.
\label{A48}
\end{equation}
An example for a unit cell is shown in Fig. \ref{fig7} with the same ratio $\xi/\gamma$ 
and hence the same shape of the small and large kinks as in Figs.~\ref{fig3},\ref{fig4}.

Summarizing, we have investigated a 3-parameter generalization of the U(1) chirally symmetric GN model. The two dimensionless
parameters $\gamma$ and $\xi$ stem from two different mechanisms of breaking chiral symmetry explicitly, the bare mass term and 
the difference between scalar and pseudoscalar couplings. Close to the chiral limit, the leading order derivative expansion has 
revealed the following scenario. If the scalar coupling dominates, we find in general a unique vacuum with scalar condensate,
light pions and kink-like baryons with baryon number 1.  In the region $\xi>\gamma$ a false vacuum shows up in the form of a second
local minimum. If the pseudoscalar coupling dominates, at first nothing changes. Starting from a critical strength of the coupling
($\xi<-\gamma$), two symmetric minima appear together with scalar and pseudoscalar condensates; parity is spontaneously
broken. The mechanical model of a particle on a rotating circle in the gravitational field illustrates nicely the concomitant pitchfork bifurcation.
The two ways of connecting two minima along the chiral circle are reflected in two baryons whose baryon numbers add up to 1.
These chirally twisted baryons are mathematically well known from studies of the double sine-Gordon equation and quite different from
another type of twisted bound state specific for the chiral limit \cite{16,17}. In our case, the baryons are stabilized by topology. Shei's bound 
state
is stabilized by partially filling the valence level and does not carry baryon number as a result of a cancellation with induced fermion
number \cite{15}. In many respects the limits $\gamma\to 0$ and $\xi\to 0$ are atypical so that previously explored 2-parameter versions
of the present
model cannot convey the full picture of chiral symmetry breaking in 4-fermion models. In view of the rich structure of the 3-parameter model, 
it seems worthwhile to pursue its study, in particular to explore the fate of the symmetries at finite temperature and chemical potential.

\section*{Acknowledgement}

This work has been supported in part by the DFG under grant TH 842/1-1.

\end{document}